\documentclass[12pt] {article}
\usepackage{amstex}
\usepackage{epsfig}

\def\ZZ{\hbox{\it Z\hskip -4.pt Z}}
\def\CC{\hbox{\it l\hskip -5.5pt C\/}}
\def\RR{\hbox{\it I\hskip -2.pt R }}
\def\MC{\hbox{\it I\hskip -2.pt M \hskip -7 pt I \hskip - 3.pt \CC}_n}
\newcommand{\nc}{\newcommand}
\nc{\beq}{\begin{equation}}
\nc{\eeq}{\end{equation}}
\nc{\beqa}{\begin{eqnarray}}
\nc{\eeqa}{\end{eqnarray}}
\def\agt{
\mathrel{\raise.3ex\hbox{$>$}\mkern-14mu\lower0.6ex\hbox{$\sim$}}
}
\def\alt{
\mathrel{\raise.3ex\hbox{$<$}\mkern-14mu\lower0.6ex\hbox{$\sim$}}
}
\begin {document}

\bibliographystyle{unsrt}    
\pagestyle{empty}
\Large
{\centerline{\bf Extension of sine-Gordon field theory}}
\vspace{3mm}
{\centerline{\bf from generalized Clifford algebras.}}
\vspace{2mm}

\large

\vspace{7mm}

\centerline {\bf P.~Baseilhac and P.~Grang\'e}

\small\normalsize
\vspace{3mm}

\centerline {Laboratoire de Physique Math\'ematique, Universit\'e Montpellier II}

\vspace{3mm}

\centerline {Place E.~Bataillon, 34095 Montpellier, France}

\vspace{7mm}

\large

\centerline {\bf M.~Rausch de Traubenberg}

\small\normalsize

\vspace{3mm}

\centerline {Laboratoire de Physique Th\'eorique}

\vspace{3mm}

\centerline {3 rue de l'Universit\'e, 67084 Strasbourg, France}

\begin{abstract}
Linearization of homogeneous polynomials of degree $n$ and $k$ variables leads
to generalized Clifford algebras. Multicomplex numbers are then introduced in 
analogy to complex numbers with respect to usual Clifford algebra. In turn 
multicomplex extensions of trigonometric functions are constructed in terms of
`compact' and `non-compact' variables. It gives rise to the natural extension
of the $d-$dimensional sine-Gordon field theory in the $n-$dimensional 
multicomplex space. In dimension 2, the cases $n=1,2,3,4$ are identified as 
the quantum integrable Liouville, sine-Gordon and known deformed Toda models. The general case is discussed.  
\end{abstract}

\vspace{10mm}
\centerline{\large Mod. Phys. Lett. {\bf A 13} (1998) 2531}
\vspace{7mm}

\vspace{1mm} 
\newpage
\pagestyle{plain}

\setcounter{page}{1}

In most domain of research activities apparently disparate facts and data are 
first accumulated and later on classified and interpreted in terms of an 
underlying hidden order. There are numerous examples in physics ranging from 
Maxwell's early unification of electricity and magnetism to the actual 
understanding of particle physics and elementary  interactions. The hidden 
structure takes the form of underlying algebras. A recent celebrated example 
relates to the interpretation of the universality of critical phenomena in 
two dimensions in term of Virasoro algebra. Among the possible algebras those 
induced by {\it bilinear} relations have a special status, probably due to the 
bilinear aspects of fundamental objects such as quadratic metric, commutators,
 anticommutators, etc$\cdots$.  However, algebras going beyond the quadratics 
ones have been constructed in the 70's from underlying polynomials of degree 
higher than two. There are dubbed by mathematicians as Clifford algebras of 
polynomials \cite{cp,cp2}. The matrix representations of such algebras 
\cite{line} lead to natural algebraic extensions of the Clifford and Grassmann
 algebras \cite{gca,gca2}. These families endow extension of complex 
numbers \cite{mc1,mc2} leading to the corresponding extension of 
trigonometric functions, dubbed multisine functions (see also \cite{hdr}).

The potential usefulness of the multisine functions has already been pointed out and explored only briefly so far. In the context of field theory it is natural to focus on one of the `simplest' (apparently) famous form, the sine-Gordon  (SG) model, known for its integrability and fermionization properties related to the massive Thirring model for a specific relation between the couplings \cite{Col,Zamo,Kla}.\\  By analogy, Toda and affine Toda field theories (TFT and ATFT) based on simply or non-simply laced algebras are the natural simple Lie group extension of the $SU(2)$ and $\hat{SU}(2)$ Toda and affine Toda field theories, {\it e.g.} Liouville and SG models. The interest in such deformed Toda theories comes from their integrability and duality properties \cite{integrable ATFT and duality,ATFTd2} which can be used to pave the way to the understanding of electric-magnetic duality in four-dimensional gauge theories, conjectured in \cite{elec-mag duality1} and developed in 
\cite{elec-mag duality2}. Thereby non-perturbative analysis of the spectrum 
and of phase structure in 
supersymmetric Yang-Mills theory becomes possible. 

Although the filiation of these models is recognized in the way already mentionned there is at present no common framework to link their studies. It is the purpose of this letter to point out such a possibility as a consequence of the properties of underlying generalized Clifford algebras. 

The mathematical formalism is first introduced. A representation 
of the multisine functions is constructed on the basis of `compact' and 
`non-compact' directions. Two sets of variables appear: one set, namely 
[$\phi_p$], represents the $E(n/2)$ `compact' directions and the other, 
namely [$\varphi_p$], represents the $n-1-E(n/2)$ `non-compact' directions
(where $E(x)$ stands for the integer part of $x$). 
Clearly for $n=2$ only one compact direction exists corresponding to the 
standard trigonometric angle. Then, the explicit expressions of the multisine 
are obtained for all $n$.

We constuct next the natural extensions of the sine-Gordon model with 
the substitution of the 2-dimensional complex space by the $n-$dimensional
 multicomplex space. The multicomplex extension of order $n$ of the cosine function appears naturally hence the name of multisine-Gordon model (MSG).

In two dimensions and for specific values of the coupling constants, the first
 cases $n=1,2,3,4$ are clearly identified as integrable quantum field theories (IQFT), $n=1$ corresponding to $sl(2)$-TFT (Liouville), $n=2$ to $sl(2)$-ATFT (SG), while $n=3$ and $n=4$ to $sl(2)$-ATFT deformed respectively by one and two fermionic interactions of the $\overline{\psi}\psi$ type. These two cases are related respectively to integrable deformations of the non-linear sigma model in Witten's black hole metric and of the $N=2$ supersymmetric SG.

Finally we envisage the cases $n=5$ and 6. The integrability of the two models is discussed. Some conclusions and perspectives are finally drawn.\\

The set of multicomplex (MC) numbers is  generated by one element $e$ 
which satisfies $e^n=-1$ \cite{mc1} 
($\MC=\left\{x= \sum \limits_{i=0}^{n-1} x_i e^i,
         e^n=-1, x_i \in \RR\right\}$),
and constitute a $n-$dimensional commutative algebra. It is worth stressing 
that most of the results of usual complex mumbers analysis remains true for MC-numbers, were it be for algebraic \cite{mc1} or analytic properties \cite{mc2}.

Having defined an appropriate pseudo-norm  ($\|x\|$) \cite{mc1, mc2}, it is possible to  write any  multicomplex number $x$ ($\|x\| \neq 0$) in an exponential representation $x=\sum \limits_{i=0}^{n-1}x_i e^i = \rho \exp{\left(\sum \limits_{i=1}^{n-1} \Theta_i e^i\right)}$. Of course due to the multivaluedness of the logarithm function the $\Theta$'s are not unique. Then, it becomes  straightforward to define the extension of the usual trigonometric functions \cite{mc1}
$x=\rho \exp{\left(\sum \limits_{i=1}^{n-1} e^i \Theta_i\right)}=
\rho \sum \limits_{i=0}^{n-1} \mathrm{mus}_i(\Theta) e^i$.

These functions have properties analogous to the usual sine and cosine functions
\cite{mc1}. However, the parametrisation in terms of $\Theta$ given in \cite{mc1,mc2} is inconvenient to derive explicit formulas for the mus-functions. From the properties of $e$ in its matrix representation (see \cite{mc1}) it is easy to see that $e^i+e^{n-i}$ (resp. $e^i-e^{n-i}$) are represented by pure imaginary (real) matrices, and, hence among the $n-1$ directions in $\sum \limits_{i=1}^{n-1} \Theta_i e^i$ one can extract $E(n/2)$ compact directions and $n-1-E(n/2)$ non-compact ones. Performing the change of basis  (for $n$ even) 
$\sum \limits_{i=1}^{n-1} \Theta_i e^i = \sum \limits_{a=0}^{n/2-1} 
\Big[\phi_a p_a + \varphi_a (-p_a^2) \Big]$, with
$p_a$ ($a=0, \cdots n/2-1$) given by $p_a= {2 \over n} \sum \limits_{j=0}^{n-1} \sin\Big[(2a+1) j {\pi \over n}\Big]e^j$, we get 
\beqa 
\label{mus2}
\mathrm{mus}_k(\phi_a,\varphi_a) &=& 
{2 \over n} \sum \limits_{a=0}^{n/2-1} 
\cos \Big[\phi_a -{(2a+1) k \pi \over n} \Big] \exp\left(\varphi_a\right),\ \ \mbox{for} \ \ k=0,\cdots n-1, \label{museven} \\
&&{\mathrm{~with~~}} \varphi_0 + \varphi_1 + \cdots + \varphi_{n/2-1} = 0. \nonumber
\eeqa

This last equation is necessary to ensure unimodular MC-numbers
and to match the number of $\Theta$'s with the number of
$\phi$ and $\varphi$'s. This
constraint of unimodularity translates into a $n-$th order relation among the
mus-functions \cite{mc1}.

In the odd case, the situation is more intricate, and the appropriate change
of basis is $e^i, i=1,\cdots n-1 \to p_a,-p^2_a -2 d, a=0,\cdots (n-1)/2-1$ with $p_a$ given as before and $d={ 1 \over (n-1)/2} \sum \limits_{i=1}^{(n-1)/2} (-1)^i e^i$,  leading to
\beqa 
\label{mus3}
\mathrm{mus}_k(\phi_a,\varphi_a) &=& 
{2 \over n} \sum \limits_{a=0}^{(n-1)/2-1} 
\cos \Big[\phi_a -{(2a+1) k \pi \over n} \Big] \exp\left(\varphi_a\right)\label{musodd} \\ 
&+&
{(-1)^k \over n} \exp{(\tilde \varphi)},\ \ \mbox{for} \ \ k=0,\cdots n-1, \nonumber \\
&&{\mathrm{~with~~}} \tilde \varphi=-2 \sum \limits_{a=0}^{(n-1)/2-1} 
\varphi_a. \nonumber
\eeqa

As for $n$ even the constraint among the non-compact variables translates into a
$n-$th order constraint among the mus-functions.\\

Now, let us study the embedding of known models in multicomplex extensions. To construct some new models of field theory as extensions of the sine-Gordon model, the first obvious step is to consider potential terms as multisine functions. Although conceptually simple it is surprisingly fruitful in relating apparently different field theoretic models. We introduce a field theory associated to the $n-$dimensional multicomplex space by : 
\begin{eqnarray}
\label{MSG}
{\cal{A}}_k^{(n)} = \int d^dx \Big[\frac{1}{2}(\partial_{\mu}\Phi)^2 + J_k 
\mathrm{mus}_k(\alpha_a\phi_a,\beta_b\varphi_b)\Big],\label{action}
\end{eqnarray}
where $J_k$ is real and ($\alpha_a$,$\beta_b$) are real coupling constant.
\begin{itemize}
\item $\Phi =(\phi_0,..,\phi_{\frac{n}{2}-1},\varphi_1,..,
\varphi_{\frac{n}{2}-1})$ if $n$ is even,
\item $\Phi =(\phi_0,..,\phi_{\frac{n-1}{2}-1},\varphi_0,..,
\varphi_{\frac{n-1}{2}-1})$ if $n$ is odd.
\end{itemize}

One may restrict to the case $k=0$ since appropriate translations in the 
compact directions generate multisine functions of higher orders from 
$\mathrm {mus}_0$ \cite{mc1}. Then, in the sequel, we only consider (\ref{MSG}) with $k=0$  ($J_0=-\mu$). Furthermore, we choose $d=2$. At classical level these Lagrangians are unbounded from below except for the cases $n=1$ and $n=2$. However we shall consider the quantum versions which becomes properly defined due to the presence, upon fermionization, of appropriate counterterms.

In both cases ($n$ even or odd), the QFT (\ref{action}) can be cut in two parts. One part can be viewed as $\frac{n}{2}-1$ (resp. $\frac{n-1}{2}$) conformal field theories (CFTs) $a=1,...,\frac{n}{2}-1$ (resp. $a=0,...,\frac{n-1}{2}-1$):
\begin{eqnarray}
{\cal{A}}^{a}_{CFT} = \int d^2x \Big[\frac{1}{2}(\partial_{\mu}\phi_a)^2 + \frac{1}{2}(\partial_{\mu}\varphi_a)^2 - \frac{2}{n}\mu e^{\beta_a\varphi_a}\cos(\alpha_a\phi_a)\Big], 
\end{eqnarray}
with an holomorphic stress-energy tensor ($z=x_1+ix_2$, $\overline{z}=x_1-ix_2$ are the complex coordinates) :
\begin{eqnarray}
T_a(z) = -2\pi\Big[(\partial_z \varphi_a)^2 + (\partial_z \phi_a)^2 - 4Q_{\beta_a}\partial_z^2\varphi_a\Big]
\end{eqnarray}
with $Q_{\beta_a} = \frac{1}{4\beta_a}$, $\alpha_a^2 - \beta_a^2 = 4\pi$ and central charge $c_a=2+\frac{3(8\pi)}{2\beta_a^2}$. For $n$ even, we add one other CFT (for $a=0$) which corresponds to a free field theory (with central charge $c_0=1$). The other part (the last term of eqs. (\ref{museven}) and (\ref{musodd})) corresponds to a perturbation which destroys conformal invariance and generates masses.

By comparison for ATFT the addition of the extra root term destroys in the same manner the TFT conformal invariance. Of course, we could write these QFT in a way similar to ATFT (defining a root system, labels, etc...) but it does not seem to give more insight. 

Except for the cases $n=1$ and $n=2$ (Liouville and sine-Gordon models) we 
do not know any Lax pair with a vanishing commutator for the classical equations of motion which could proof the classical integrability of these theories. Moreover some analog to the Cartan-Weyl basis, useful for Toda models, is not obvious here. 

On the other hand, if a theory is quantum integrable in (1+1) dimensions, the 
conservation of currents can be obtained from massless perturbation theory in 
which the whole exponentials terms are treated as interaction terms. This is 
essentially equivalent to OPE techniques. As for massive two dimensional field
 theories, the ultraviolet divergent diagrams are tadpoles, removed by 
normal-ordering\,\footnote{The normal-ordered version of vertex operator is : $\exp(\beta \phi): = \exp(\beta \phi) /a^2$ where $a$ is a cut-off. However, this multiplicative constant leads to no modification regarding OPE calculations. }.  We work in Euclidian space, with $(z,\overline{z})$ 
coordinates. The massless bosonic propagators are identical and such that \ 
$<\phi(z,\overline{z})\phi(0,0)>= <\varphi(z,\overline{z})\varphi(0,0)>=-
\frac{1}{4\pi}\ln(2z\overline{z})$ . Local holomorphic currents (with spin $s$) of the form ($n$ even):
\begin{eqnarray}
T^{(s)}&=& \sum_{r_i,r'_i} X_{p_1...p_{\frac{n}{2}-1},q_0...q_{\frac{n}{2}-1},
p'_1...p'_{\frac{n}{2}-1},q'_1...q'_{\frac{n}{2}-1}}^{r_0...
r_{\frac{n}{2}-1},r'_1...r'_{\frac{n}{2}-1}} (\partial^{p_0}
\phi_{q_0})^{r_0}...(\partial^{p_{\frac{n}{2}-1}}
\phi_{q_{\frac{n}{2}-1}})^{r_{\frac{n}{2}-1}} \\ \nonumber 
&&\ \ \ \ \ \ \ \ \ \ \ \ \times \ \ \ \ (\partial^{p'_1}
\varphi_{q'_1})^{r'_1}...(\partial^{p'_{\frac{n}{2}-1}}
\varphi_{q'_{\frac{n}{2}-1}})^{r'_{\frac{n}{2}-1}},\nonumber
\end{eqnarray}
and their anti-holomorphic counterpart\,\footnote{Anti-holomorphic currents can be defined as in \cite{qc}.} $\overline{T}^{(s)}$ are then considered, with $\partial=\frac{\partial}{\partial z}$, \ \ $\sum  p_i r_i + p'_i r'_i  = s$\ and \ 
$X_{p_1...p_{\frac{n}{2}},q_1...q_{\frac{n}{2}},p'_1...
p'_{\frac{n}{2}-1},q'_1...q'_{\frac{n}{2}-1}}^{r_1...
r_{\frac{n}{2}},r'_1...r'_{\frac{n}{2}-1}}$ determined - up to the shift 
$T^{(s)} \rightarrow T^{(s)} + \partial{\Lambda^{(s)}}$ - by the
conservation laws :
\begin{eqnarray}
\overline{\partial} <T^{(s)}{\cal{V}}(\phi,\varphi)> + \partial <\overline{{T}}^{(s)}{\cal{V}}(\phi,\varphi)> =0,
\end{eqnarray}
where the expectation value is taken on the Fock space vacuum of the massless quantum field theory and ${\cal{V}}(\phi,\varphi)$ stands for the different vertex operators which appears in the multisine potential. The coefficients $X_{p_0...p_{\frac{n}{2}-1},q_0...q_{\frac{n}{2}-1},p'_1...p'_{\frac{n}{2}-1},q'_1...q'_{\frac{n}{2}-1}}^{r_0...r_{\frac{n}{2}-1},r'_1...r'_{\frac{n}{2}-1}}$  possess the same symmetry properties as the Lagrangian, resulting in a set of strong constraints among them.
Generalization to the case $n$ odd is straightforward, with a change 
$\frac{n}{2}-1$ to $\frac{n-1}{2}$.\\
The $X$'s are then determined in a way such as to cancel contributions coming 
from potential anomalies - e.g. local contributions which cannot be written 
as the $\partial$-derivatives of some suitable expression.
\begin{itemize}
\item Case $n=1$ : the Liouville model.
\end{itemize}
In this case, the multicomplex space is just the real space. 
Strickly speaking there is no exponential representation in the
sense given before. However, if instead of unimodular numbers,
positive numbers are used, one may write $x=\exp{(\varphi)}$ leading to the 
potential $\exp(-2\beta\varphi)$. This field theory is then integrable as 
it corresponds to a CFT, the well-known Liouville model : 
\begin{eqnarray}
{\cal{A}}_0^{(1)} = \int d^2x \Big[\frac{1}{2}(\partial_{\mu} \varphi)^2 - \mu 
\exp(-2\beta\varphi)\Big].\label{L}
\end{eqnarray} 
\begin{itemize}
\item Case $n=2$ : the sine-Gordon model.
\end{itemize}
For $n=2$ we immediately obtain the well-known sine-Gordon model \cite{Col,Kla} 
which is integrable \cite{Zamo} for $d=2$:
\begin{eqnarray}
{\cal{A}}_0^{(2)} = \int d^2x \Big[\frac{1}{2}(\partial_{\mu} \phi)^2 - 
\mu\cos(\alpha\phi)\Big].\label{SG}
\end{eqnarray}
The 2-d fermion-boson correspondence \cite{Col} gives :
\begin{eqnarray}
{\cal{A}}_0^{(2)} = \int d^2x \left(i\overline{\psi} \gamma_{\mu}
\partial_{\mu}\psi -\frac{g}{2}(\overline{\psi}\gamma_{\mu}\psi)^2 - 
\mu\overline{\psi}\psi\right),\label{Thirring}
\end{eqnarray}
with : 
\begin{eqnarray}
g/\pi = \frac{4\pi}{\alpha^2}-1.\label{g}
\end{eqnarray}
\begin{itemize}
\item Case $n=3$.
\end{itemize}
Starting from the expression (\ref{action}) for $n=3$ the action writes :
\begin{eqnarray}
{\cal{A}}_0^{(3)} = \int d^2x \Big[\frac{1}{2}(\partial_{\mu} \Phi)^2 - 
\frac{\mu}{3} \big(( 2\exp(\beta_0\varphi_0)\cos(\alpha_0\phi_0)+
\exp(-2\beta_0\varphi_0) \big)\Big].
\label{n=3}
\end{eqnarray}
This model is integrable \cite{Fateev} for :
\begin{eqnarray}
\alpha_0^2-\beta_0^2=4\pi, \label{condition}
\end{eqnarray}
and the integrals $Q_2$ of spin 2 are generated by the conserved currents $T_3$ of spin 3 which have the form :
\begin{eqnarray}
T_3 = A(\partial\phi_0)^3\ +\ B(\partial\phi_0)(\partial\varphi_0)^2\ + \ 
C(\partial^2\phi_0)(\partial\varphi_0),
\end{eqnarray}
with : $A = 2b^2+a^2,\ B = 3b^2,\ C = -3b(1+4b^2)$\ and 
$a=\frac{\alpha_0}{\sqrt{8\pi}}$, $b=\frac{\beta_0}{\sqrt{8\pi}}$.

For $\mu=\frac{3}{2}(\frac{M^4}{\beta_0^2})^{\frac{1}{3}}$ and with the shift $\varphi_0 \rightarrow \varphi_0-\frac{1}{3\beta_0}\ln(\frac{\sqrt{M}}{\beta_0^2})$  this QFT possesses \cite{Fateev} a dual representation corresponding to the complex sinh-Gordon (CSG), which is a sigma model with Witten's black hole metric. For small $\beta_0$, one can use the 2-d fermion-boson correspondence which gives :
\begin{eqnarray}
{\cal{A}}_0^{(3)} = \int d^2x \Big[\frac{1}{2}(\partial_{\mu} \varphi_0)^2 + 
i\overline{\psi} \gamma_{\mu}\partial_{\mu}\psi -\frac{g}{2}(\overline{\psi}
\gamma_{\mu}\psi)^2 &-& M \exp(\beta_0\varphi_0)\overline{\psi}\psi -\frac{M^2}{2\beta_0^2}\exp(-2\beta_0\varphi_0)\nonumber \\
&+& CT^{(3)}(\varphi_0,\beta_0)\Big],
\label{n=3'}
\end{eqnarray}
with g defined by (\ref{g}) and $CT^{(3)}(\varphi_0,\beta_0)=-\frac{M^2}{2\beta_0^2}\exp(2\beta_0\varphi_0)$ a usual counterterm added to cancel the fermion loop divergence.  Furthermore, it has the $U(1)$ symmetry generated by the charge $Q = \int dx \overline{\psi}\gamma_{0}\psi$. The mass spectrum\,\footnote{The boson $\varphi_0$ is unstable and does not appear in the spectrum.} is made of fermions ($\psi$, $\psi^{\dagger}$) with mass $M$ \cite{Fateev}. 

\begin{itemize}
\item Case $n=4$.
\end{itemize}
Starting from the expression (\ref{action}) for $n=4$, the following action 
results 
\begin{eqnarray}
{\cal{A}}_0^{(4)}&=& \int d^2x \Big[\frac{1}{2} (\partial_{\mu} \varphi_1)^2 +
 \frac{1}{2} (\partial_{\mu} \phi_0)^2 + \frac{1}{2} (\partial_{\mu} 
\phi_1)^2 \\ \nonumber
&& \ \  \ \ \ \ -\frac{\mu}{2} \left( \exp(-\beta_1\varphi_1)
\cos(\alpha_0\phi_0) + \exp(\beta_1\varphi_1)\cos(\alpha_1\phi_1) 
\right)\Big] .\label{QFT3}
\end{eqnarray}
Restricting to $\alpha_0 = \alpha_1= \alpha$ and $\beta_1=\beta$, the model is integrable if 
(\ref{condition}) is verified \cite{Fateev}(with the substitution $\alpha_0 \rightarrow \alpha$ and $\beta_0 \rightarrow \beta$).\\ 
The first non-trivial spin 4 current $T_4$ generating the conserved charge 
$Q_3$ is :
\begin{eqnarray}
T_4 &=& A\Big[(\partial\phi_0)^4 + (\partial\phi_1)^4\Big] + B(\partial\varphi_1)^4 + C\Big[(\partial\phi_0)^2(\partial\phi_1)^2 + (\partial\phi_0)^2(\partial\varphi_1)^2 + (\partial\phi_1)^2(\partial\varphi_1)^2\Big] \nonumber \\
&+& D\Big[(\partial^2\varphi_1)(\partial\phi_0)^2 - (\partial^2\varphi_1)
(\partial\phi_1)^2\Big] + E\Big[(\partial^2\phi_0)^2 + (\partial^2\phi_1)^2\Big] + F(\partial^2\varphi_1)^2,
\end{eqnarray}
with : $A = a^2(1+3b^2),\ B= \frac{b^2}{2}(3a^2-1),\ C = 6a^2b^2,\ D= 6a^2b,\ 
E = 2(1-6a^2+10a^4),\\ F = 1-11a^2+7a^4$ and $a=\frac{\alpha}{\sqrt{8\pi}}$ ; $b=\frac{\beta}{\sqrt{8\pi}}$.

Moreover, the QFT (\ref{QFT3}) has $U(1)\times U(1)$ symmetry generated by 
the charges $Q_a = \int dx \overline{\psi_a}\gamma_{0}\psi_a$. 
For small $\beta$, the 2-d fermion-boson correspondence gives :
\begin{eqnarray}
{\cal{A}}_0^{(4)}&=& \int d^2x \Big[\frac{1}{2} (\partial_{\mu} \varphi_1)^2 +
 \sum_{a=0}^{a=1}\left( i\overline{\psi_a} \gamma_{\mu}\partial_{\mu}\psi_a -
\frac{g}{2}(\overline{\psi_a}\gamma_{\mu}\psi_a)^2\right) \\ \nonumber
&& \ \  \ \ \ \ -\frac{\mu}{2} \sum_{a=0}^{a=1}\left( \exp(\beta_a\varphi_a)
\overline{\psi_a}\psi_a \right) + {CT}^{(4)}(\varphi_1,\beta)\Big] ,
\end{eqnarray}
with $g$ defined by :
\begin{eqnarray}
g/\pi = \frac{4\pi}{\alpha^2}-1 = -\frac{\beta^2}{\alpha^2} = -
\frac{\frac{\beta^2}{4\pi}}{1 + \frac{\beta^2}{4\pi}},\label{g_4}
\end{eqnarray}
 $\beta_0=1$ and $CT^{(4)}(\varphi_1,\beta) \sim \cosh(2\beta\varphi_1)$.\\
For $\mu=2M$ and $CT^{(4)}(\varphi_1,\beta)=-\frac{M^2}{2\beta^2}\cosh(2\beta\varphi_1)$ this QFT was studied in \cite{Fateev} and is related to $N=2$ supersymmetric 
SG  theory, anisotropic principal chiral field, the O(4) non-linear sigma 
model,  etc... The factorized scattering theory gives the S matrix as a 
product of  two SG S-matrices. 
\begin{itemize}
\item Other cases.
\end{itemize}

The cases $n=5$\,\footnote{It can be checked that these cases correspond to $C^{(1)}_2$ and $D^{(2)}_2$-ATFT deformed by two fermionic interactions \cite{mus2}} and $n=6$ initiate an ensemble ($n$ even, $n$ odd) of models 
possessing  similar vertex operator structure. In keeping with the earlier 
cases, more generally for $\beta_a$ small and for $k=$0, the $2-d$ 
fermion-boson correspondence \cite{Col,Kla} can be used to rewrite the action (\ref{action}) in a form more suitable for perturbative analysis, using for $d=2$ the definitions (\ref{action}). We can define also $CT^{(n)}(\varphi,\beta)$, the appropriate counterterms to cancel the divergences coming from the fermions loops.  For example the counterterms of $n=3$ and $n=4$ correspond respectively to the Liouville and sinh-Gordon potential ($SU(2)$ and $\hat{SU}(2)$ Toda potential). These two series of QFT possess $U(1) \times ... \times U(1)$ symmetry generated by the charges $Q_a = \int dx \overline{\psi_a}\gamma_{0}\psi_a$. The quantum integrability of these higher order QFT is an open question. There are some arguments pointing to the 
integrability of the case $n=5$ using algebra embeddings. Indeed, the
mus-fuctions we have considered, as well as there $n-$order associated 
constraint, are basically related to the matrix realization of the MC-algebra
\cite{mc1}. However, it is also possible to find a matrix representation
of $\MC$ of dimension higher than $n$. This is essentially related to
the embeddings
$\hbox{\it I\hskip -2.pt M \hskip -7 pt I \hskip - 3.pt \CC}_n \subset
\hbox{\it I\hskip -2.pt M \hskip -7 pt I \hskip - 3.pt \CC}_m$ when
$n < m$ \cite{mc1} 
(in fact to be more precise we have to distinguish odd and even
$n$, but this subtility is irelevant here). With such an embedding we can
obtain other mus-fuctions of order $n$. The expression of the
mus are similar to those given in  (\ref{mus2}-\ref{mus3}) but an other constraint is obtained in the non-compact direction. This translates into a relation of degree $m$ on the mus instead of a relation of degree 
$n < m$. Detailed work is in progress \cite{mus2} and will be reported elsewhere. So far our investigations indicate that the cases $n>5$ seem more problematic.

Finally, some remarks are necessary about the parameter $\mu$ introduced which is very important as it defines the mass spectrum obtained for bosonic fields $\varphi_a$. Then, the study of dynamics is more problematic as Yang-Baxter equations becomes more complicated and non-diagonal scattering may appear. However, integrability still holds.
\begin{itemize}
\item Concluding remarks.
\end{itemize}

The concept of unifying algebras has long proved fruitful in the 
understanding of apparently disparate physical phenomena. Major advances and 
developments in critical phenomena in two dimensions occured from properties
 of local conformal invariance with the underlying Virasoro algebra. 
Quantum field theory in two dimensions has accumulated a variety of integrable
 models ranging from Liouville, sine-Gordon to Toda and affine Toda theories, 
etc...In TFT the coupling constant is chosen real or imaginary and a similar transformation leads to the equivalent extension of sinh-Gordon theory, opening of a new domain of investigation however beyond the scope of this exploratory letter. 
In the search for a common framework we have shown that generalized Clifford algebras and their associated multicomplex structure and trigonometry provide a strikingly unifying scheme for a whole set of QFT models of higher 
and higher complexities. The main characteristics of the primary models is 
their known integrability. It is believed that this new framework should 
provide some insight on the properties of QFT's of higher order in the 
hierarchy as well as to lead to some genuine developments in models of 
statistical physics.

Finally to conclude, we would like to mention that the specific algebraic
structure of the MC-numbers should be relevant for a general analysis.
One can compare for instance the property of the model $n=4$ studied where the matrix $S$ can be written as a product of two SG $S-$matrix with the property that $\hbox{\it I\hskip -2.pt M \hskip -7 pt I \hskip - 3.pt \CC}_4$ contains the set of complex number \cite{mc1}. Anyhow, if we use higher dimensional representations for the MC-number, we  have, in principle, other models than those described in (\ref{action}) with  (\ref{mus2}-\ref{mus3}), having other constraints upon the non-compact variables.

\paragraph*{Aknowledgements}

V. A. Fateev and A. Neveu are gratefully aknowledged for useful discussions.

\end{document}